\documentclass[preprint]{revtex4-1}
\usepackage{amsmath}
\usepackage{graphicx}
\usepackage{natbib}
\usepackage{lineno}

\draft % marks overfull lines with a black rule on the right

\begin{document}
\title{Molecular simulation of surface reorganization and wetting in crystalline cellulose I and II}
\author{Reinhard J. Maurer}
\affiliation{Department Chemistry, Technical University Munich, Lichtenbergstr. 4, D-85747 Garching, Germany}
\author{Alexander F. Sax}
\email[]{alexander.sax@uni-graz.at}
\author{Volker Ribitsch}
\affiliation{Department of Chemistry, University of Graz, Heinrichstr. 28, A-8010 Graz, Austria}


\begin{abstract} Cellulose is one of the most versatile
substances in the world. Its immense variety of applications
was in recent years complemented by nanotechnological
applications such as cellulose nanoparticle dressed surfaces
for filtration purposes or cellulose matrices for
microelectronics. The fabrication of such complex materials
asks  for thorough understanding of the surface structure and
its interactions with adsorbates. In this study we investigate
several surface model systems of nanotechnological interest,
which are obtained by reorganization of the cellulose-vacuum or cellulose-water interfaces
of slabs of crystalline cellulose. To do this, we equilibrated first bulk supercells of
different cellulose allomorphs, which were constructed from
crystallographic data, and then optimized the interface structures.
From the bulk and surface systems we calculated
structural properties such as unit cell parameters, dihedral
conformation distributions, density profiles and hydrogen
bonding. The results suggest that no overall geometrical restructuring
occurs at the interface. However, the hydrogen bond network is
strongly reconstructed, as is inferred from the dihedral conformations
and hydrogen bond occurrences, although only within the first few layers. This
holds for low index close packed structures as well as for high index
loosely packed surfaces. Replacing the vacuum by ambient pressure water molecules
we find less rearrangements of the cellulose surface, because
the water allows formation of hydrogen bonds similar to those
in the bulk phase. The water near the cellulose surface shows,
however, strong structural changes. We observe reduced mobility
of the water molecules, which corresponds to a cooling of water by about 30 degrees,
 in a slab that is  about 10\,\AA\ thick. Although structuring
and adsorption is observed on all surfaces, no actual penetration of water
into the cellulose structure could be observed. This suggests that pure water
is not sufficient to produce cellulose swelling at mesoscopic timescales. This work
lays the basis for current quantum chemical investigations on specific interaction terms
within cellulose.
\end{abstract}

\maketitle

\section{\label{intro} Introduction}

Cellulose is one of the most abundant materials on earth
\citep{Sullivan97}.
Exploitation of the vast amount of this renewable substance
dates back until ancient times. The recent uprising of
nanotechnology has opened a new range of applications for
cellulose and its derivatives. It is used for cellulose
nanocomposites, such as flexible, transparent matrices for
organic light emitting diodes (OLED) \citep{Eichhorn10}, or
stable cellulose nanoparticles \citep{Hornig08}. To obtain new
smart materials the surface properties of crystalline cellulose
must be properly modified. The importance of this branch of
carbohydrate research has been underlined by approval of a
research proposal within the European Union 7th framework
program, called SURFUNCELL, wherein 15 academic and industrial
partners try to formulate nanostructured cellulose coatings
with a variety of proposed applications, \textit{e.g.} self
cleaning surfaces or waste water treatment.

A systematic formulation of technologically interesting
materials such as the one planned in the mentioned project
needs deep understanding of the underlying molecular structure
of and reactivity at surfaces of the solid phases. This is,
however, complicate because cellulose is a polymorph with
several crystalline forms such as I$\alpha$, I$\beta$, II, III,
but also amorphous forms \citep{Sullivan97}, which are partly
coexistent. Cellulose is built from $\beta (1\rightarrow 4)$
glucan molecules which are linear chains of D-glucose monomers linked
by glycosidic bonds. In the two allomorphs I$\alpha$ and I$\beta$ of native
cellulose all chains are parallel aligned having the same
growth direction, the allomorphs differ in their crystal
structures: I$\alpha$ has a triclinic unit cell of $P1$ symmetry with
a single cellobiosyl unit as basis, the crystal consists of
parallel layers of one type, I$\beta$ has a monoclinic unit cell of
$P2_1$ symmetry with two cellobiosyl units as basis, termed
origin and center, accordingly the crystal is made of two
different alternating sheets.
Regenerated cellulose II has a monoclinic unit cell of $P2_1$
symmetry with two cellobiosyl units. \citep{Kolpak76}
Cellulose III is obtained when either cellulose I or cellulose II is treated with
liquid ammonia, the resulting allomorphs are consequently denoted as
III$_1$ or III$_2$, both allomorphs are believed to have $P2_1$
unit cells. \citep{Wada2004} This transformation is reversible.
Heating of cellulose III leads to cellulose IV$_1$ and
IV$_2$, respectively, which can reversibly be transformed into
cellulose I and II, respectively, both allomorphs have a $P_1$ unit cell.

The fact that native cellulose is composed of crystalline and
amorphous domains makes the structural investigation very complicated.
Nonetheless, many different X-ray structures of cellulose were
published over the years \citep{Gardner74a, Gardner74b,
Kolpak76, Langan99, Nishiyama02, Nishiyama03}. These reports
gave a very fundamental understanding of the molecular
structure, but many structural aspects are inaccessible by
X-ray diffraction, such as the hydrogen bond (HB)  network or the
structure of cellulose surfaces or interfaces between different
phases.

Molecular modeling and especially molecular dynamics (MD)
represent reliable tools to investigate structural and
dynamical properties based on preliminary experimental results.
Considering polysaccharides \citep{Perez96} and in particular
cellulose many different modeling approaches are presented in
literature, reaching from studies of bulk properties
\citep{Perez96, Kroon-Batenburg97, Mazeau03, Vietor00,
Heiner95}, thermal response \citep{Bergenstrahle07}, relative
stability \citep{Kroon-Batenburg96, Hardy96} and deformation
\citep{Eichhorn05}, to studies of solvated crystallites
\citep{Matthews06, Yui06, Yui07}, surfaces properties or
interaction with water\citep{Heiner97, Biermann01, Mazeau2008,
Bergenstrahle08} and adsorbates\citep{Mazeau02}.

Films of regenerated cellulose are semicrystalline or amorphous,
that means that highly ordered crystalline domains of cellulose II
coexist with disordered domains within their elementary supramolecular
units. Crystalline and amorphous cellulose differ in many physical and
chemical aspects,  amorphous cellulose is, for example, accessible to water,
in contrast to crystalline cellulose.\citep{Kontturi11} Since most chemistry
with cellulose is done in aqueous solutions, understanding of surface processes
like adsorption needs a detailed knowledge of properties of the cellulose-water
interface, however, little data on the crystalline ordering and crystal
structure of cellulose films are available.\citep{Aulin09}
Modeling of the heterogenous surface structure of cellulose films is, therefore,
rather difficult, it is more difficult for the amorphous than for the crystalline
domains. When the crystalline domains are treated as crystallites the
surfaces are best described by crystal planes; modeling of amorphous domains that
are accessible to water can in principle be done with very large supercells and
very irregular surfaces, but there are no data available, which can be used for
modeling these surfaces. It is assumed that initial solubilization processes with water
are only possible when many polar groups are exposed at the surfaces and when the surfaces area is maximal.
In order to investigate surface reorganization of cellulose we choose surfaces
generated from large Miller index crystal planes and compare them with
low index surfaces. We specifically address the reconstruction
of the cellulose/vacuum and cellulose/water interfaces and the influence of the
surface morphology on the properties of the water component.

\section{\label{methods} Methods}

Throughout this study, molecular dynamics (MD) and optimization
runs were performed with the DL\_POLY 2.20 package
\citep{Smith05} using periodic boundary conditions. We used the
all atom force field GLYCAM06 (Glycam06f.dat)\citep{Kirschner08} to
describe the cellulose crystal systems and a non-rigid TIP3P
model \citep{Jorgensen83} to describe the water molecules. For
the system setup we used AmberTools 11 \citep{Case05} and
ChemShell 3.3.1 \citep{Sherwood03}, therewith producing DL\_POLY
readable force field files. Manipulations on the molecules were
done using the model builder Aten \citep{Youngs10}, and
visualization with VMD 1.8.6 \citep{Humphrey96}. Calculations
were performed on an AMD64 Linux 24 node cluster with double
precision.

\begin{figure}
\includegraphics[width=\textwidth]{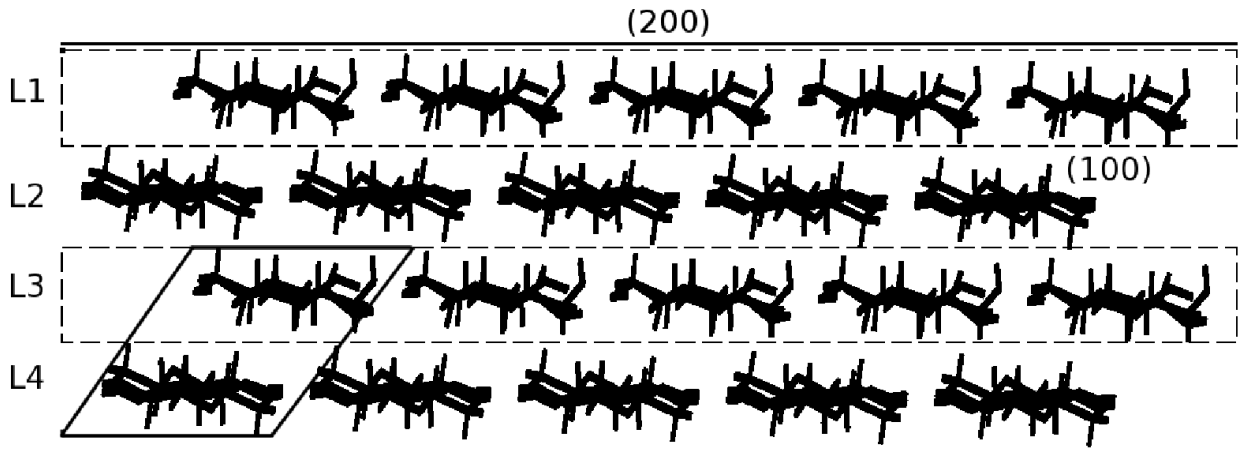}
\caption{\label{fgr: layer} Schematic representation of a
cellulose I$\beta$ surface.  The small parallelogram at the
left bottom depicts the unit cell of the system. The sheets are
separated by dashed lines and defined as slabs of cellulose
chains orthogonal to the envisaged surface (represented by the
black line at the top). In this case the sheets defined by the
elementary cell also act as layers in the slab.}
\end{figure}

All surface models are derived from the respective optimized
bulk models, therefore, we first optimized the bulk properties
of the investigated allomorphs. Each bulk model is represented
by supercells which are repeats of the respective unit cells.
The I$\alpha$ supercell has $5\times 4 \times 5$ repeats of the
triclinic unit cell and the I$\beta$ supercell has $2\times 5 \times
5$ repeats of the monoclinic unit cell along the corresponding
$\vec a$, $\vec b$ and $\vec c$ directions. The cellulose II
(010) surface model is based on a bulk model which has $5\times
2 \times 5$ repeats of the monoclinic unit cell, termed IIa.
Unit cell parameters and coordinates of the cellobiose units
were taken from recent X-ray diffraction
studies\citep{Nishiyama02, Nishiyama03, Langan99}. All three
systems contained 4200 atoms and their volume was about
33$\;$nm$^3$. The cell parameters are the respective X-ray
values in Table 1. For the cellulose II (120) surface model we
constructed a second bulk supercell IIb with $2\times 2 \times
4$ repeats of a unit cell with 8 cellobiose units ( a =
29.27$\;$\,\AA, b = 8.93$\;$\,\AA, 10.36$\;$\,\AA, $\alpha$ = $\beta$
= 90$^\circ$, $\gamma$ = 80.15$^\circ$) 5376 atoms in a volume
of 44$\;$nm$^3$. (\textit{cf.} Fig. \ref{fgr: layer})

These systems were equilibrated and optimized in sequences of
molecular dynamics and geometry optimizations, similar to a
simulation protocol of \citep{Mazeau03}: 20$\;$ps NVT MD at
600$\;$K, geometry optimization, 20$\;$ps NPT MD at 450$\;$K,
geometry optimization, 200$\;$ps NPT MD at 300$\;$K, geometry
optimization. For the bulk models of cellulose II this sequence
was preceded by a 40$\;$ps N$\sigma$T MD at 300$\;K$ and
geometry optimization.

The surface models are represented by slabs which consist of
the corresponding optimized bulk supercells augmented by  a
vacuum slab of the same size. To model the I$\alpha$ (100),
I$\beta$ (100), II (010) and the II (120) surfaces we
thus doubled the size of our bulk supercells in one direction
which was the b, a, b and b direction, respectively. All
modeled systems consist of 4 sheets of cellulose chains
parallel to the  respective surfaces, with the lowest cellulose
sheet being frozen to simulate a bulk boundary. To find out,
whether this choice of slab is a too severe size limitation we
performed some test calculations on cellulose I$\beta$ and IIa
with up to 8 sheets but did not find significant differences
with respect to the measured observables, therefore the system
size was restricted to 4 sheets throughout this study.

The surface/vacuum systems were equilibrated by the sequence:
20$\;$ps NVT MD at 600$\;$K, geometry optimization, 200$\;$ps
NPT MD at 300$\;$K, geometry optimization.

For the surface/water systems we started from the respective
optimized surface/vacuum systems where the vacuum slab was
replaced by a  22$\;$\,\AA\ thick layer of water (density
$\rho$ = 1$\;$g$\,$cm$^{-3}$). The layer of water was built using the
tool Packmol\citep{Martinez09}. The systems were treated as
follows: geometry optimization, 500$\;$ps NVT MD at 300$\;$K,
geometry optimization.

All molecular dynamics runs were performed using a 1$\;$fs time
step, Ewald summation and Nose-Hoover thermostats and barostats
with relaxation times $\tau_t$ = 0.1$\;$ps and $\tau_p$ =
1$\;$ps. Observed quantities for bulk and surface systems were
averaged over the last 100$\;$ps of the corresponding 300$\;$K MD run.

The self-diffusion coefficients for water were calculated using the Einstein relation.
The mean-squared displacement data were averaged over the last 200$\;$ps of the
trajectory. The short averaging time does not allow an error estimate, but averaging
over a significantly longer run did not show any qualitative changes to the results.

\begin{figure}
\includegraphics[width=0.6\textwidth]{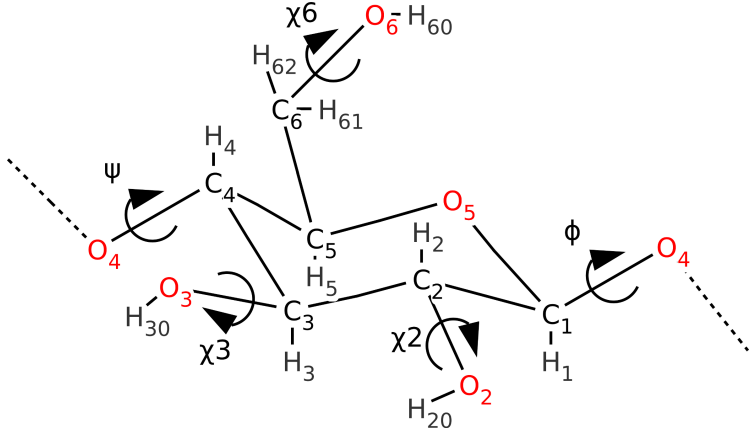}
\caption{\label{fgr: naming} GLYCAM06 naming convention used in this work. }
\end{figure}

In this work we use the GLYCAM06 \citep{Kirschner08}
naming convention for glucose residues (Fig. \ref{fgr: naming}).

\section{\label{results}Results and discussion}

\subsection*{\label{bulk} Bulk systems }

Starting from the experimental structures obtained from X-ray diffraction,
we equilibrated the bulk systems according to the
scheme described in section \ref{methods}. Only few of the relaxed
cell parameters listed in Table \ref{tbl:unit cell} differ considerably from  the
experimental cell parameters. We find an increase of the cell
volume by about 5\% with respect to the X-ray structures, which
is mainly caused by the much larger c parameter, whereas the a
and b parameters as well as the cell angles are very close to
the experimental starting values. The reason for the larger c
parameters is probably the underestimation of the puckering of
the glucose rings by the force field resulting in flatter pyranose rings
and, therefore, longer glucan chain. This seems to be a
general problem of force fields as comparison with the results
of other modeling studies for cellulose I$\beta$ shows
(see Table \ref{tbl:unit cell}).

\begin{table}
  \caption{Unit cell parameter of N$\sigma$T equilibrated systems (this study), and values from the literature. Experimental studies are marked with Exp.}
  \label{tbl:unit cell}
  \begin{tabular}{ccccccccc}
    \hline
    Crystal phase & Study & a [\,\AA] & b [\,\AA] & c [\,\AA] & $\alpha$ [deg]& $\beta$  [deg]& $\gamma$  [deg]& V[\,\AA$^3$] \\
    \hline
    Cellulose I$\alpha$ & This Study & 6.89(3) & 5.83(3) & 10.75(1) & 120.4(4) & 110.6(4) & 81.7(3) & 348.3 \\
                & \citep{Nishiyama03} (Exp) & 6.717 & 5.962 & 10.400 & 118.08 & 114.8 & 80.37 & 333.3   \\
    Cellulose I$\beta$ & This Study & 7.90(3) & 8.40(1) & 10.75(1) & 85.1(2) & 90.2(4) & 102.2(5) & 694.6 \\
                & \citep{Nishiyama02} (Exp) & 7.784 & 8.201 & 10.380 & 90 & 90 & 96.5 & 658.3  \\
                & \citep{Mazeau03} & 8.378 & 8.168 & 10.523 & 89.97 & 89.97 & 90.92 & 720.0  \\
                & \citep{Matthews06} & 8.47 & 8.11 & 10.51 & 90 & 90 & 90 & 721.9 \\
                & \citep{Zhang11}, 298 K    & 7.63 & 8.23 & 10.80 & 89.99 & 89.99 & 97.17 &  672.9\\
                & \citep{Zhang11}, 500 K    & 8.11 & 8.28 & 10.78 & 90.00 & 89.96 & 98.33 &  716.2  \\
    Cellulose IIa & This Study & 8.02 & 9.32 & 10.77 & 90.2 & 91.6 & 121.5 & 686.1 \\
                & \citep{Kolpak76} (Exp) & 8.01 & 9.04 & 10.36 & 90 & 90 & 117.1 & 667.8  \\
                & \citep{Langan99} (Exp) & 8.10 & 9.05 & 10.31 & 90 & 90 & 117.1 & 672.8 \\
   Cellulose IIb  & This Study & 8.30 & 8.77 & 10.73 & 89.7 & 91.2 & 117.1 & 694.1 \\
    \hline
 \end{tabular}
\end{table}

A very recent study by Zhang \emph{et al.}\citep{Zhang11} investigated,
using MD simulation in the NPT ensemble and the GLYCAM06
force field, the structural response of  crystalline  I$\beta$ to heating from room
temperature to 600 K. Experimentally, the change from a low temperature structure
to a high temperature structure is found to be reversible, but in their MD simulation
the phase transition was irreversible. The parameters of the unit cell for both
structures in Table \ref{tbl:unit cell}
show mainly an increase in the $a$ parameter and a slight increase of $\gamma$,
giving an increase of the volume of the unit cell.

Chen \textit{et al.} \citep{Chen12} found the same irreversible change from the low to the high
temperature structure in a simulation with the united atoms force field GROMOS 53a6
and attributed this fault to the parameters of the torsion potential for the exocyclic
hydroxymethyl group. With properly changed potential parameters the phase transition
became reversible.

There is no full agreement between our cell parameters for I$\beta$ and those
by Zhang \textit{et al.}, but this is not surprising, because Zhang \textit{et al.}  systematically
heated the system from room temperature to 500 K, whereas in the mixture of MD runs
and geometry optimization described above for the equilibration of the bulk systems
the highest temperature was 600 K. Moreover, the super cell we used had different
size and shape. Our $a$ parameter is in-between the corresponding
low and high temperature values of Zhang \textit{et al.}, the $c$ parameters agree very
well as do the $\beta$ values. Our $b$ value is closer to the high temperature value,
our $\alpha$ value is by 5 degrees smaller than both values by Zhang \textit{et al.}, and our $\gamma$
value is 4 and 5 degrees larger than the high and low temperature values by Zhang \textit{et al.},
the volume of our elementary cell is in-between the low and high temperature  values.
We assume that the equilibrated bulk cell obtained with our scheme is neither a full
low temperature nor a full high temperature structure.

\begin{figure}
\includegraphics[width=0.6\textwidth]{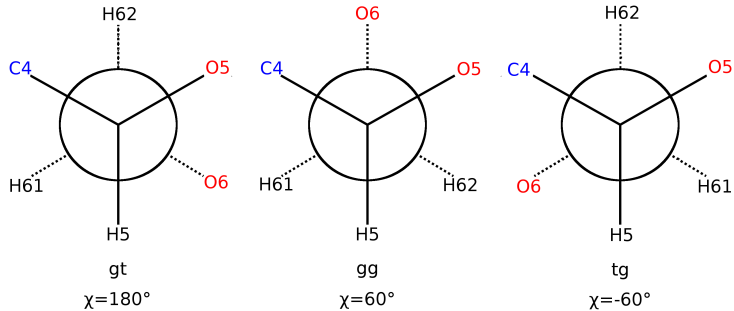}
\caption{\label{fgr: C6dih} Different conformations of C6 group.
 $\chi$ is defined as C4-C5-C6-O6 dihedral angle. }
\end{figure}

The phase transition is accompanied by a  change in the conformation of the exocyclic
hydroxymethyl group, which can be found in three staggered conformations termed
\emph{gg}, \emph{gt} and \emph{tg}  where the first \emph{g} and \emph{t} mean gauche
and trans position, respectively, of the C6-O6 group with respect to C5-O5, whereas
the second \emph{g} and \emph{t} mean gauche and trans position of the C6-O6 group with
respect to C5-C4. The conformations can be characterized by, e. g., the values of the C4-C5-C6-O6
dihedral angle $\chi$ (Figure \ref{fgr: C6dih}).  In the \emph{tg} and \emph{gt}
conformations the position of the C6-O6 bond is equatorial with respect to the glucose six ring,
in the \emph{gg} conformation the C6-O6 bond is axial. In the \emph{tg} conformation can accordingly
the O6H group participate in intra- and inter-chain hydrogen bonds,
in the \emph{gt} conformation it participates only in interchain hydrogen bonds and in the \emph{gg} conformation
the O6H group  can participate only in inter-sheet hydrogen bonds.
Zhang \textit{et al.}  found that upon heating $tg$ conformations in the origin sheets are dominantly changed
into $gt$ conformations, but into $gg$ conformations in the center sheets.\citep{Zhang11}
Chen \textit{et al.} \citep{Chen12} showed that above 500 K the \emph{tg} conformations
are completely replaced by \emph{gt} and \emph{gg} conformations.

\begin{figure}
\includegraphics[width=0.5\textwidth]{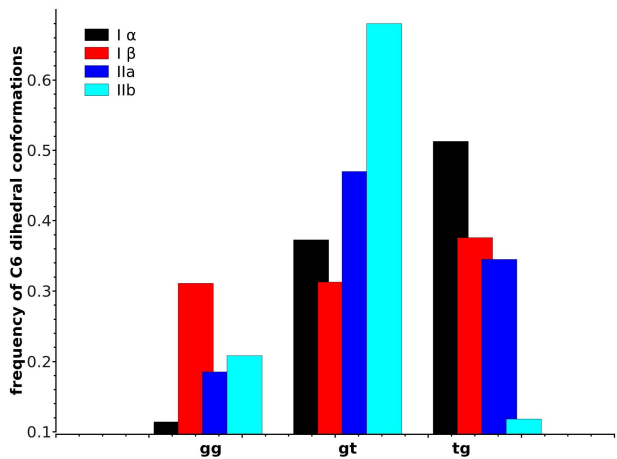}
\caption{\label{fgr: C6BULK} Frequency of the hydroxymethyl conformations.}
\end{figure}

Figure \ref{fgr: C6BULK} shows the distribution of the hydroxymethyl conformations in the 4 cellulose systems investigated.
In I$\beta$  the ration of \emph{tg} : \emph{gt} : \emph{gg} is $38:31:31$, so our I$\beta$ system is neither the low
temperature system with 100 percent \emph{tg}, nor is it the high temperature system with no \emph{tg} conformations at all.
The equal population of  \emph{gt} and \emph{gg} conformations is in accord with the findings of Zhang \textit{et al.} and
Chen \textit{et al.}
Considering the influence of the force field on the population of the hydroxymethyl conformations, one can assume that the
high \emph{gt} population in I$\alpha$ is also an artefact of the force field, the extremely low \emph{gg} population is in
agreement with  experimental findings.

For the investigation of cellulose II we used two different supercells, the distribution of the hydroxymethyl conformations
differs considerably for the  \emph{gt} and \emph{tg} conformations: in both systems  \emph{tg} dominates but in IIa the
percentage is below 50 percent and in IIb it is about 66 percent. In both systems the \emph{gg} population is about 20 percent.
We assume  that the difference is caused by the equilibration scheme and the different size and shape of the
supercells.

\subsubsection*{Hydrogen bonds in cellulose bulk}
A hydrogen bond (HB) is a system where a hydrogen atom is covalently bonded to a donor oxygen OD-H bond, which  weakly interacts
with the lone pair electrons of an acceptor oxygen atom OA, that is not directly bonded to the group.  Such a HB is
represented by ODH...OA. Since the OH bond lengths in different HBs are rather constant, the non-bonded O...H distance and the
OH...O angle are sufficient to characterize a HB.

Two types of intra-chain HBs between glucose rings can be found in a glucan molecule;
the first one is O3-H...O5 with the  hydroxyl group O3H in one glucose ring
and the acceptor atom O5 in the next glucose ring; the second   HB is O2-H...O6 with the hydroxyl group
O2-H in one ring and the acceptor atom O6 in the second ring. This HB is only possible when the exocyclic
hydroxymethyl group has \emph{tg} conformation.
Thus, the covalent glycosidic bond and the two HBs build a strong link between two glucose rings
and make the chain rather stiff. Interchain HBs connect glucan molecules to sheets, which interchain HBs are formed depends
on the crystal structure and on the temperature.

Table  \ref{tbl: Ialfabulk} shows the relative occurrence of the different HBs in the equilibrated I$\alpha$ bulk.

Most important are the intra-chain HB O3H...O5 and the inter-chain HB O2H...O6; with 32 percent each, they represent
nearly two third of the hydrogen bond network in the bulk, to the remaining third contribute the intra-chain HB
 O6H...O2 HB and  the O6H...O3 inter-chain HB with about 16 percent each. That means, about half of the
 hydroxymethyl groups are  found in the \textit{tg} and the other half in the \textit{gt} conformation.
 The rest of the HBs are weak intra-chain and inter-sheet HBs.

\begin{table}
\caption{ Distribution of hydrogen bonds in I$\alpha$ bulk together with  HB parameters. Only hydrogen bonds  with a
percentage higher than 0.5 are shown. Distances are given in \AA ngstrom, angles in degree.
}
  \label{tbl: Ialfabulk}
  \begin{tabular}{ccrrr}
    \hline
    HB & type$^a$ & rel. occ. & $d(\rm O\cdots HO)$ & $\angle(\rm OHO)$\\
    \hline
    O5$\cdots$HO3 & A &  32 & 1.86 & 155 \\
    O6$\cdots$HO2 & E &  32 & 1.88 & 153 \\
    O6$\cdots$HO3 & E &   1 & 2.06 & 159 \\
    O3$\cdots$HO6 & E &  16 & 1.91 & 157 \\
    O2$\cdots$HO6 & A &  16 & 1.93 & 159 \\
    O2$\cdots$HO3 & A &   1 & 2.09 & 146 \\
    O4$\cdots$HO6 & S &   1 & 2.01 & 153 \\\hline
  \end{tabular}
\\
$a$) A means intra- and E means inter-chain, and S means inter-sheet hydrogen bonds.
\end{table}

\begin{figure}[ht]
\includegraphics[width=\textwidth]{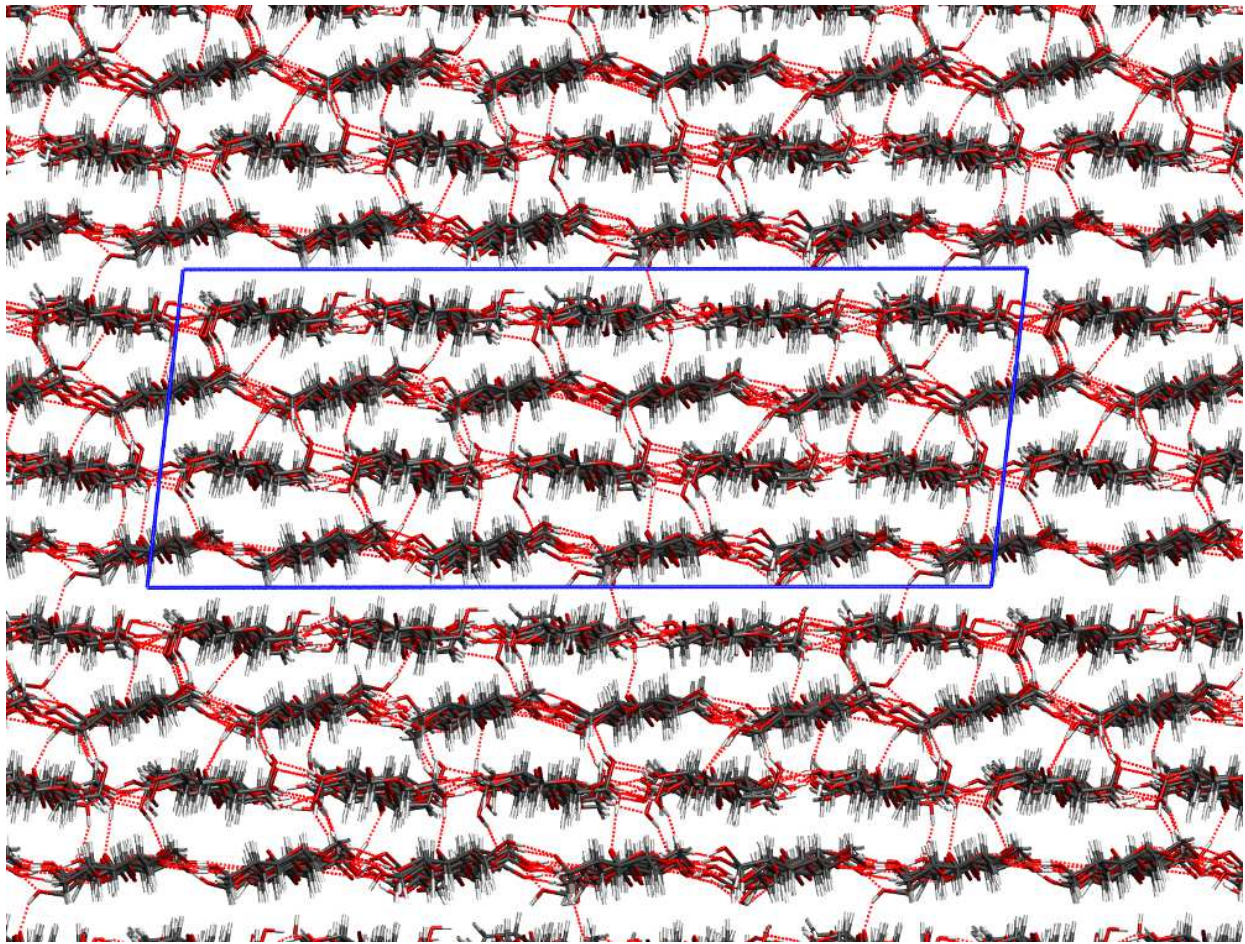}
\caption{\label{fgr: Ibetaslab} a,b base plane projection of the cellulose I$\beta$ (100) slab. }
\end{figure}

Crystalline cellulose I$\beta$  is made of two alternating sheets
(Figure \ref{fgr: Ibetaslab}) corresponding to the two different cellobiosyl units in the  unit cell, called center and origin
units. In one sheet the chains are nearly coplanar, in the other sheet  the chains
are slanted, Heiner \textit{et al.} \citep{Heiner97,Heiner98} called these sheets ''odd'' and ''even'', we prefer ''flat'' and
''rippled''. According to authors like  Nishiyama \textit{et al.} \citep{Nishiyama03} the origin units create the flat sheets
others like Zhang \textit{et al.}\citep{Zhang11}  make the flat sheet with the center units.
 Nishiyama \textit{et al.} \citep{Nishiyama02} showed for the first time the existence of different
HB networks in the two sheets, Figure \ref{fgr: Ibetaslab} visualizes the different HB pattern.

The analysis of the HB network in equilibrated I$\beta$ bulk shows that in the rippled sheets there are about 20 percent
more HBs than in the flat sheets, which corroborates the existence of two different HB networks. A detailed comparison of the
two networks reveals equal numbers of intra-chain HBs  O3H..O5 and inter-chain HBs O2H..O6 in both sheet types,  these
two HBs form the equal backbone of the HB networks. The difference between the two sheet types comes from the different
numbers of O6H..O3 inter-chain HBs, which is in the rippled sheets about 2 to 3 times higher than in the flat sheets. That
means, the stiffer chains in the flat sheets are connected mainly by O2H..O6 inter-chain HBs, the less stiff chains in the
rippled sheets are connected by the same number of inter-chain HBs O2H..O6 and by additional O6H..O3 inter-chain HBs.

\begin{table}
\caption{ Distribution of HBs in I$\beta$ bulk together with  HB parameters. Only bonding atoms  with a
percentage higher than 0.5 are shown.
}
  \label{tbl: Ibetabulk}
  \begin{tabular}{crrr}
    \hline
    HB & rel. occ. & $d(\rm O\cdots HO)$ & $\angle(\rm OHO)$\\
    \hline
      & \multicolumn{3}{c}{bulk}\\
    O3H$\cdots$O5 &  34 & 1.83 & 157 \\
    O2H$\cdots$O6 &  34 & 1.91 & 158 \\
    O3H$\cdots$O6 &   0 & 2.07 & 148 \\
    O6H$\cdots$O3 &  14 & 1.99 & 154 \\
    O6H$\cdots$O2 &  12 & 2.00 & 154 \\
    O6H$\cdots$O4 &   6 & 1.96 & 157 \\\hline
    & \multicolumn{3}{c}{flat sheets}\\
    O3H$\cdots$O5   &  42 & 1.83 & 156\\
    O2H$\cdots$O6   &  40 & 1.96 & 156\\
    O3H$\cdots$O6   &   0 & 2.01 & 150\\
    O6H$\cdots$O3   &  11 & 2.05 & 157\\
    O6H$\cdots$O2   &   7 & 2.01 & 155\\
    O6H$\cdots$O4   &   0 & 2.09 & 138\\\hline
    & \multicolumn{3}{c}{rippled sheets}\\
    O3H$\cdots$O5   &   35 & 1.83 & 157\\
    O2H$\cdots$O6   &   34 & 1.91 & 161\\
    O3H$\cdots$O6   &    1 & 2.04 & 149\\
    O6H$\cdots$O3   &   23 & 1.96 & 152\\
    O6H$\cdots$O2   &    8 & 2.14 & 155\\
    O6H$\cdots$O4   &    0 & 2.03 & 137 \\ \hline
  \end{tabular}
\end{table}
Compared with I$\alpha$, the number of \textit{tg} conformations of the hydroxymethyl group decreases in both sheet types
similarly strong, but the increase of \textit{gt} and \textit{gg} conformations in the two sheet types is very
different:
in flat sheets there are more \emph{gg} conformations than \textit{gt} conformations, in rippled sheets there are nearly no
\textit{gg} conformations.
Consequently, most hydroxymethyl groups in rippled sheets are involved in inter-chain HBs, whereas in flat sheets
they are also involved in  inter-sheet HBs. This was also found by Zhang \textit{et al.}\citep{Zhang11}.
Relative frequencies and geometry parameters of the HBs are given in Table \ref{tbl: Ibetabulk}.

This result diverges considerably from the  C13 CP/MAS results that the hydroxymethyl group in I$\alpha$ and I$\beta$ has
only the \emph{tg} conformation.\citep{Wada2004_1} Experimentally, two different hydrogen bond networks exist in I$\beta$, but in both networks the hydroxymethyl group has only the \emph{tg} conformation.

\begin{figure}[ht]
\includegraphics[width=\textwidth]{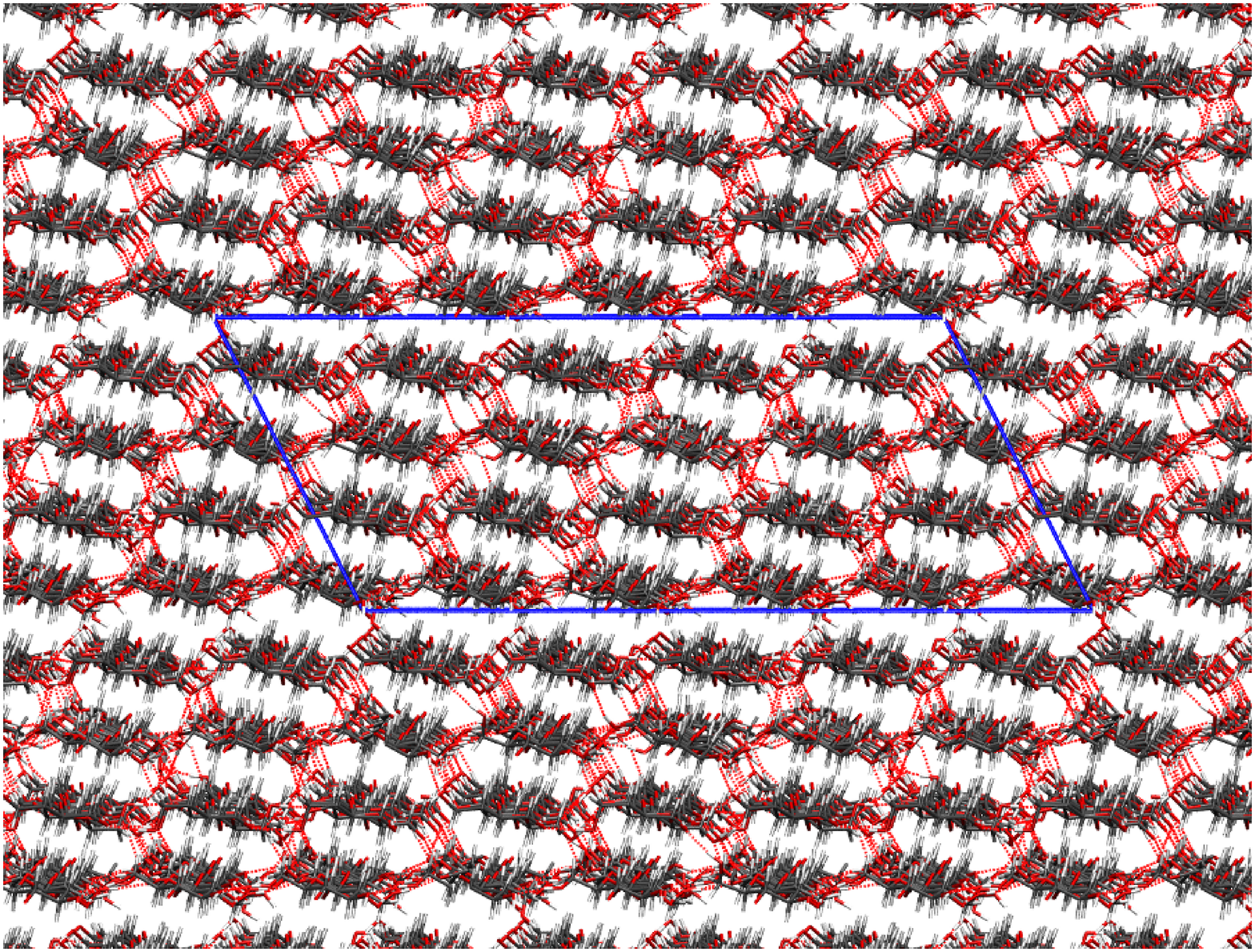}
\caption{\label{fgr: IIaslab} a,b base plane projection of the cellulose II bulk supercell. }
\end{figure}

In analogy with cellulose I$\beta$, monoclinic cellulose II crystals are made of alternating  sheets, but the growth
directions of the glucan molecules are reversed in alternating sheets.
This allows HBs which do not occur in I$\beta$ and, thus, gives rise to different topologies of the HB network.
Inspection of the a,b base plane projection of the cellulose II supercell in figure \ref{fgr: IIaslab} shows clearly the
existence of two sheet types but a distinction between flat and rippled seems to be rather artificial. In the a,b base plane
projection of the origin sheets, which are made of the origin units in the unit cell, one can observe
a cyclic structure of the inter-chain HBs (sheets 1 and 3 in Figure \ref{fgr: IIaslab}), we find mainly
O6H..O2 and little O6H..O3 HBs, the structure is highly regular as nearly all O6H groups are involved in these HBs;
the alternating position
of the glucose rings that are connected by the inter-chain HBs causes the regular HB network. That means also that in origin
sheets there are no O6H..O2 intra-chain HBs; the intra-chain HB O3H...O5 has the same frequency as in I$\beta$.

For the center sheets, which are made of the center units, no cyclic structure of the inter-chain HBs can be seen in
the a,b base plane projection. We observe a low regularity in the HB network, which is mainly made from O2H..O6 and O6H..O3
inter-chain HBs, O2H...O6 HBs occur nearly exclusively in the center sheet. Not all O6H..O2 intra-chain HBs are broken and
replaced by inter-chain HBs, but the number of O3H...O5 intra-chain HBs is in the center sheets only half of that in the origin
sheets, the O3H groups are mainly targets of  O6H groups. In center sheets one can also find O2H..O6 intra-chain HBs.

Inter-sheet interaction mainly happens via O2H..O2 HBs with the O2H group from the origin sheet, and O6H..O2 HBs, with the O6H
group from the center sheet, but there are also O3H..O6 and O6H..O6 inter-sheet HBs.

\begin{table}
 \caption{ Distribution of HBs in IIa bulk together with  HB parameters. Only bonding atoms  with a
 percentage higher than 0.5 are shown.
 }
  \label{tbl: IIabulk}
  \begin{tabular}{crrr}
    \hline
    HB & rel. occ. & $d(\rm O\cdots HO)$ & $\angle(\rm OHO)$\\
    \hline
      & \multicolumn{3}{c}{bulk}\\
    \hline
    O3H$\cdots$O5 &  23 & 1.94 & 153.6 \\
    O2H$\cdots$O6 &  17 & 1.82 & 161.3 \\
    O6H$\cdots$O3 &  17 & 1.85 & 156.2 \\
    O6H$\cdots$O2 &  14 & 1.83 & 158.8 \\
    O3H$\cdots$O6 &  10 & 2.03 & 153.0 \\
    O6H$\cdots$O6 &       &       &       \\
    O6H$\cdots$O4 &       &       &       \\
    O2H$\cdots$O2 &  17 & 1.85 & 2.830\\  \hline
        & \multicolumn{3}{c}{origin sheet}\\
    O3H$\cdots$O5 &  47 & 1.93 & 155.4\\
    O2H$\cdots$O6 &   1 & 1.98 & 154.9\\
    O6H$\cdots$O3 &  12 & 1.94 & 160.5\\
    O6H$\cdots$O2 &  39 & 1.82 & 158.9\\
    O3H$\cdots$O6 &   1 & 1.82 & 157.6\\
    O6H$\cdots$O6 &   1 & 1.98 & 154.9\\
    O6H$\cdots$O4 &       &       &      \\\hline
        & \multicolumn{3}{c}{center sheet}\\
    O3H$\cdots$O5 &  21 & 1.96 & 149.4\\
    O2H$\cdots$O6 &  47 & 1.81 & 161.0\\
    O6H$\cdots$O3 &  28 & 1.82 & 155.0\\
    O6H$\cdots$O2 &   1 & 2.12 & 150.0\\
    O3H$\cdots$O6 &   2 & 2.02 & 150.7\\
    O6H$\cdots$O6 &       &       &      \\
    O6H$\cdots$O4 &   0 & 1.97 & 139.3\\\hline
  \end{tabular}
\end{table}
Note, the relative occurrences for the bulk in Table \ref{tbl: IIabulk} are not the sum of the occurrences for the layers,
the latter do not contain any inter-sheet HBs.

\subsubsection*{Thermodynamic stability of cellulose bulk}
By averaging over configurational energies of the equilibrated
systems we get an estimate of the relative thermodynamic
stabilities of the different allomorphs.
From experiment\citep{Yamamoto89,Kroon-Batenburg96} it is known that
heating of cellulose I$\alpha$ will convert it into I$\beta$ and finally
into II, but not the other way around. Experimentally,
cellulose II is thermodynamically most stable and I$\alpha$ is least
stable, our simulations yield the following ordering of the energies per
cellobiose unit: 215.0\,kJ\, mol$^{-1}$ (I$\beta$), 213.8\,kJ\, mol$^{-1}$
(I$\alpha$), 211.6\,kJ\, mol$^{-1}$ (IIa) and 210.7\,kJ\, mol$^{-1}$ (IIb).

Our results corroborate the finding that cellulose II is more stable than cellulose I,
the $d(\rm O\cdots HO)$ distances are in cellulose II shorter than
in I$\beta$, indicating that the glucan molecules are closer together and this
will increase the stability because the HB  and the dispersion interaction will be stronger.
According to our quantum chemical calculations dispersion interaction between the chains
will be more important than HBs. The expected higher stability of cellulose I$\beta$ with respect to cellulose
I$\alpha$ was, however, not found. We attribute this discrepancy to the fact that our
equilibrated I$\beta$ system is indeed a less stable high temperature system.
We are sure that the difference in the stability of
cellulose IIa and IIb is  caused by the equilibration scheme.
Since we were mainly interested in the generation of surfaces of
different roughness for the study of the cellulose-water-interface we
did not investigate the origin of the difference between the two supercells.

\subsection*{\label{surf} Cellulose/vacuum and cellulose/water interfaces}

\begin{figure}
\includegraphics[width=\textwidth]{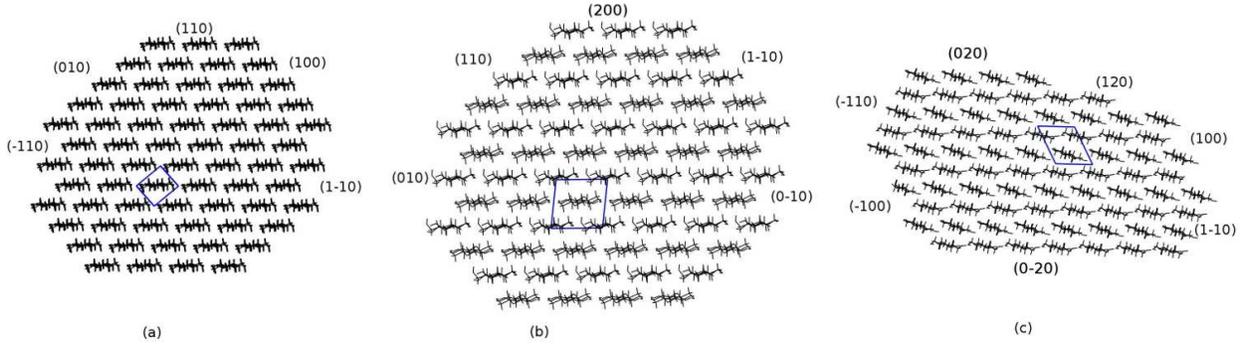}
\caption{\label{fgr: nanocrystal} Schematic structure of cellulose I$\alpha$ (a), I$\beta$ (b) and II (c) nanocrystallites, exposing different crystal planes.}
\end{figure}

The aim of this study is to investigate how the properties of the cellulose/water interface
depend on the structure of the cellulose surface. Macroscopic cellulose samples, such as fibres or films,
are often regarded as being composed of nanocrystallites embedded in amorphous cellulose;
in a first approximation, the surface of such a sample can then be seen as a tiling made of many different
crystal faces of irregular shape, which are described by the Miller indices of the corresponding crystal planes
(see Figure \ref{fgr: nanocrystal}). As models for surfaces of the amorphous parts we choose crystal planes with
large Miller indices showing large surface roughness.
The properties of the macroscopic surface is then an average of the properties of all exposed crystal faces.

It is not possible to investigate the surface properties for all exposed crystal planes, but only for few representatives which
differ most strongly in certain properties like surface roughness, polarity etc. and in the interaction with
adsorbates. This approach has also been chosen by e. g. Mazeau and Vergelati\citep{Mazeau02}. In the I$\alpha$ (100) system
the bulk sheets are not
parallel to the surface, in the layers parallel to the surface the glucose rings are slanted; in the I$\beta$ (100) the
layers are parallel to the surface and at least in the flat sheet also the glucose rings are parallel; in II (010) the
layers are
made of parallel sheets but the glucose rings are slanted;  in II (120), finally, the sheets are not parallel to the surface,
the layers are made from glucan
molecules that stem alternatively from origin and center sheets and the overall surface shows a zigzag behavior.

\begin{figure}
\includegraphics[width=0.5\textwidth]{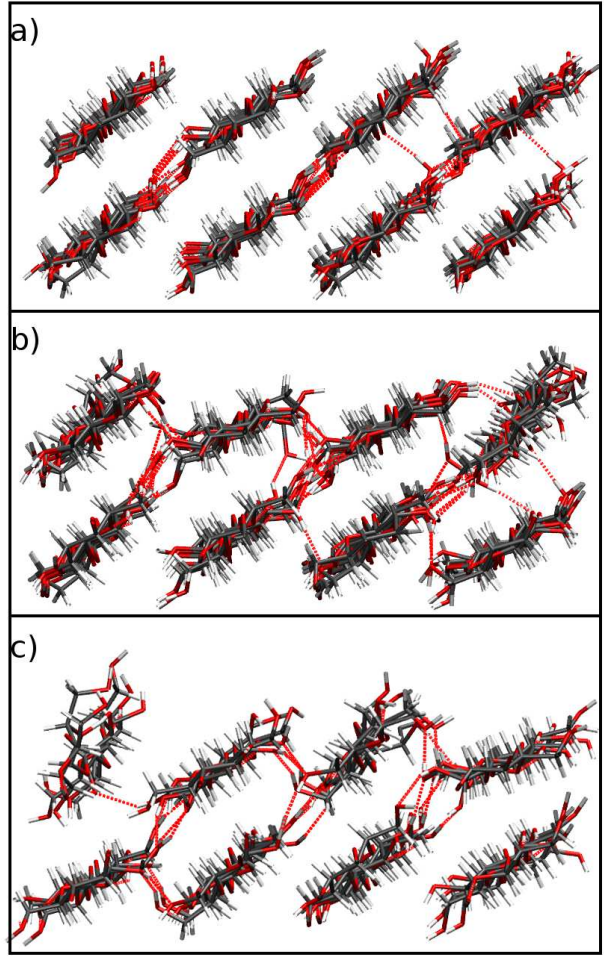}
\caption{\label{fgr:Ialpha-Reconstruction} Cutout of the supercells showing a) parts of two layers of Cellulose I$\alpha$ bulk, b) parts of the first two layers of a Cellulose I$\alpha$ (100) surface exposed to vacuum and c) parts of the first two layers of a Cellulose I$\alpha$ (100) surface exposed to 300$\,$K TIP3P water.}
\end{figure}

The aforementioned molecular dynamics scheme reveals very different degrees of reconstruction for the different surfaces.
The I$\alpha$ (100) surface, where the rings of the cellobiose molecules are slanted with respect to the (100) plane, shows
high irregularities after equilibration (shown in figure \ref{fgr:Ialpha-Reconstruction}b); inter-chain hydrogen bonds become
inter-sheet hydrogen bonds and this causes a reduction of the number of hydroxy groups pointing into the vacuum and
a strengthening of the bonding between the glucan molecules in the top layer.
Not surprising is that the flat (100)/(010) surfaces of cellulose I$\beta$ and II show little reconstruction, this could be
different for the more rippled I$\beta$ (200) surface, but we did not investigate it.
In spite of the grooved structure of the cellulose II (120) surface we did not find a considerable reconstruction due to
vacuum exposure.

To quantify the different roughnesses of the surfaces and thereby the degree of surface reconstruction we
adapted a measure which is very common in material science\citep{Degarmo03}. The R$_a$ value of a surface is
defined as follows:
\begin{equation}
R_a=\frac{1}{n}\sum_{i=1}^n |\Delta y_i| \qquad \Delta
y_i=y_i-\bar{y}
\end{equation}
Whereas y$_i$ is the absolute height of a surface atom and $\bar{y}$ defines the average surface height.
As the average surface height we took the tangential plane to the closed van der Waals spheres of the uppermost layer.

In Table \ref{tbl: roughness} we show the roughnesses of the four reconstructed surfaces. Roughness is a property of the structure of surfaces, the values from Table \ref{tbl: roughness} support the
classification of I$\beta$ (100) as smooth and of II (120) as rough, that II (010) is much smoother than I$\alpha$ (100) is not that obvious. Interestingly, the difference in roughness between a bulk layer, the surface exposed to vacuum or to a water environment is insignificant compared to the overall difference between the different cellulose types. In addition to the unchanged surface roughness we also were not able to find significant changes of the layer distances orthogonal to the surface. This suggests that at this time scale no drastic changes in the mesoscopic geometric structure of cellulose occurs.

\begin{table}
  \caption{Roughness R$_a$ [\,\AA] of cellulose surfaces exposed to vacuum.}
  \label{tbl: roughness}
  \begin{tabular}{cc}
    \hline
    Surface & R$_a$ [\,\AA] \\
    \hline
    Cellulose I$\alpha$ (100)& 0.7418 \\
    Cellulose I$\beta$ (100) & 0.5429  \\
    Cellulose II (010)       & 0.5890  \\
    Cellulose II (120)       & 0.8289  \\
    \hline
  \end{tabular}
\end{table}

Another frequently used measure for  roughness is the standard deviation of the distribution of the absolute heights
 y$_i$ from the average surface height
\begin{equation}
\tilde{R}_a=\sqrt{\frac{\sum_{i=1}^n (\Delta y_i)^2}{n-1}}
\end{equation}
The two measures give different roughness values, the $\tilde{R}_a$ values are always larger than the $R_a$ values.
For a large number of uniformly distributed random numbers $y_i$ the difference is about 16\%, for normally distributed random
numbers the difference is about 25\%. The coordinates of the top layer atoms are neither uniformly nor normally distributed random numbers, therefore, we regard the larger roughness values  $\tilde{R}_a$ reported by Mazeau\citep{Mazeau11} as consistent with our data.
%%%%%%%%%%%%%%%%%%%%%%%%%%%%%%%%%%%%%%%%%%%%%%%%%%%%%%%%%%%%%%%%%%%%%%%%%%%%%%%%%%%%%%%%%%%%%%%%%%%

Nonetheless, surface reconstruction at the solid/vacuum interface is accompanied by a significant change of the molecular hydrogen bond network, which can be
described by the frequencies of occurrence of the three conformations of the exocyclic C6 group and by the number of hydrogen bonds.
The data were obtained from the last 100\, ps of the 300\,K MD runs.

\begin{figure}[ht]
\includegraphics[width=7.5cm]{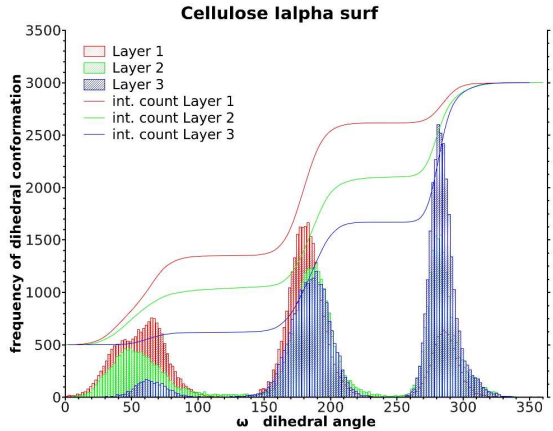}
\caption{\label{fgr: histo} Representative distributions of C6 dihedral angle conformations for different layers in a Cellulose I$\alpha$ (100) surface exposed to vacuum.
Also shown are the integrated occurrence frequencies that are used for analysis. }
\end{figure}

In Figure \ref{fgr: histo} the dihedral angles of the exocyclic C6 group are plotted which were measured for the layers in
I$\alpha$ (100), but we find for all systems similar distributions with peaks that are
easily distinguishable and can be assigned to one of the three dihedral conformations. The area under the peak can
be interpreted as the probability of finding this conformation, the relative probabilities for all systems are given in
Table \ref{tbl:C6surf+water}. The reference data for the crystalline systems with equivalent layers (I$\alpha$ and  IIb)
are obtained by averaging over all layers,
as for cellulose I$\alpha$ and  cellulose IIb, or by averaging over the two different kinds of layers, as
for cellulose I$\beta$ and  cellulose IIa. The averaged data are always given in line L3, for the systems with two different
kinds of equivalent layers, the corresponding data for L1 and L2 are also given.

In the cellulose I$\alpha$ bulk system all layers parallel to the surface show the same relative probabilities for the
dihedral conformations within small statistical fluctuations. The small amount of \emph{gg} conformations (6 percent) and the large
amount of \emph{gt} (40 percent) and \emph{tg} (54 percent) conformations once again demonstrates that in I$\alpha$ there are nearly no inter-sheet HBs, but many intra- and inter-chain HBs.
The alternating layers in the I$\beta$ system are identical with the flat and rippled sheets and have rather different
distributions of the dihedral conformations with a high frequency of  \emph{gg} conformations in the flat sheets (layers 1 and 3) and nearly no \emph{gg} conformations in the rippled sheets (layers 2 and 4), as already discussed above.

In the IIa system again the layers are identical with the sheets and we find accordingly different probability distributions
in alternating layers. In the origin sheets (layers 1 and 3) we find again a high number of \emph{gg} conformations indicating that
the C6OH groups from these sheets form active inter-sheet HBs whereas little active inter-sheet HBs are made with C6OH
groups from center sheets (layers 2 and 4).
In the IIb system the layers are made from equal contributions of origin and center glucan molecules and, therefore, all
layers show the same probability distribution. The difference between systems IIa and IIb is again demonstrated in the
averaged probability distributions which differ mostly in the \emph{gt} and \emph{tg} probabilities.

Now we compare the bulk data with the probabilities for the layers of the cellulose/vacuum interfaces.
For I$\alpha$ (100) there is a higher probability to find \emph{gg} and \emph{gt} conformation in the layers near to the surface
and a reduced probability to find \emph{tg} conformations.
The increase of \emph{gg} conformers corresponds to the increase of inter-sheet HBs between the tilted cellulose chains at
the surface. The probability distribution of layer 3 is again very close to the bulk distribution, we interpret this finding
as showing that reconstruction of the I$\alpha$ (100) surfaces is restricted to only few layers.

In cellulose I$\beta$ (100) on the other hand we find only little differences between the probability distribution for
the bulk and for the reconstructed surface, because in this system the layers are formed by sheets and the top layer is a
flat sheet. Maybe, reconstruction of the (200) surface would result in larger differences at least for the top layer,
but we did not investigate this surface.

Although in IIa (010) the layers are identical with sheets, we find strong increase in the probability for
 \emph{gg} conformations in the first two layers, the third layer shows again a distribution very similar to that in
the bulk. In IIb (120), finally, again only the two top layers are strongly reconstructed, the \emph{tg} conformations are decreased and the \emph{gt} conformations are increased.

There is, thus, a very good correspondence of surface roughness and surface reconstruction, the surface with the lowest
R$_a$ value shows the least reconstruction, irrespective of whether the layers are sheets or not. This correspondence was also
reported by Mazeau \textit{et al.}\citep{Mazeau11} for cellulose I.

When the vacuum of the solid/vacuum interface is replaced by bulk water we find very different changes of the
probability distributions: The changes in the two top layers of I$\alpha$ are much smaller than the changes between the
bulk and the solid/vacuum interface, the changes in layer 3 can be neglected. In I$\beta$, on the other hand, we see in the
top layer a strong increase of the \emph{gg} and a strong decrease for the \emph{tg} conformations, in layer 2 the \emph{gt} conformations
increase in expense of the \emph{tg} conformations. The changes in layer 3 are again small.

In IIa replacement of the vacuum by water is accompanied by little changes in the probability distributions, in IIb
a significant  increase of \emph{gt} in expense of \emph{tg} conformations is found only in the top
layer, the other changes are again rather small.

Summarizing, we found that surfaces that show strong hydrogen bonding reconstruction at the solid/vacuum interface, change little when
water replaces the vacuum and vice versa, and, none of the top layers resemble very strongly the corresponding bulk layer
hydrogen bond network, that means, there is always a surface reconstruction at an interface, the kind and amount of reconstruction of the cellulose
surface depends on the second phase at the interface, the cellulose allomorph and the structure of the crystal plane.
Similar results were obtained by Newman and Davidson\citep{Newman04} in a  $^{13}$C NMR study of the cellulose--water interfaces for cellulose I and cellulose II surfaces. From AFM studies of cellulose I$\alpha$ surfaces Baker \emph{et al.}\citep{Baker00}  conclude that hydroxymethyl groups will change at the surface from \emph{tg} to \emph{gt} conformation.

\begin{table}
  \caption{C6 dihedral angle distributions for two cellulose interfaces and crystalline bulk.
The L stands for the layer in the slab. In systems where the layers are equivalent, the average distribution
for the whole slab is shown. Details are given in the main text.}
  \label{tbl:C6surf+water}
  \begin{tabular}{ccccc}
    \hline
    Crystal system & L & crystal bulk & solid/vacuum & solid/water \\
     && {\emph{gg} : \emph{gt} : \emph{tg}} & {\emph{gg} : \emph{gt} : \emph{tg}} & {\emph{gg} : \emph{gt} : \emph{tg}} \\
    \hline
    Cellulose I$\alpha$ (100)	
                         & 1 &                    & 0.34 : 0.51 : 0.16 & 0.26 : 0.59 : 0.15 \\
		                 & 2 & 			          & 0.21 : 0.43 : 0.36 & 0.18 : 0.48 : 0.34 \\
		                 & 3 & 0.06 : 0.40 : 0.54 & 0.05 : 0.42 : 0.53 & 0.00 : 0.46 : 0.54 \\[2ex]
    Cellulose I$\beta$ (100)	
                         & 1 & 0.37 : 0.37 : 0.26 & 0.48 : 0.27 : 0.25 & 0.65 : 0.26 : 0.09 \\
		                 & 2 & 0.02 : 0.50 : 0.48 & 0.02 : 0.62 : 0.36 & 0.00 : 0.70 : 0.30 \\
		                 & 3 & 0.47 : 0.19 : 0.34 & 0.55 : 0.13 : 0.32 & 0.60 : 0.10 : 0.30 \\
		                 & 4 & 0.03 : 0.55 : 0.42 &                       &                       \\[2ex]
    Cellulose II (010) (IIa)
                     	 & 1 & 0.22 : 0.74 : 0.04 & 0.48 : 0.41 : 0.11 & 0.50 : 0.45 : 0.05 \\
  		                 & 2 & 0.08 : 0.27 : 0.65 & 0.24 : 0.34 : 0.42 & 0.16 : 0.40 : 0.44 \\
		                 & 3 & 0.21 : 0.76 : 0.03 & 0.25 : 0.71 : 0.05 & 0.24 : 0.73 : 0.03 \\
		                 & 4 & 0.06 : 0.26 : 0.68 &                       &                       \\[2ex]
    Cellulose II (120) (IIb)	
                         & 1 &                    & 0.33 : 0.41 : 0.26 & 0.34 : 0.59 : 0.08 \\
		                 & 2 &                    & 0.30 : 0.57 : 0.13 & 0.28 : 0.62 : 0.10 \\
		                 & 3 & 0.16 : 0.73 : 0.11 & 0.19 : 0.73 : 0.09 & 0.17 : 0.79 : 0.07 \\
    \hline
  \end{tabular}
\end{table}

The only significant difference between results obtained from the simulations with a 4 layer and an 8 layer supercell,
respectively, was found for the distribution of the C6 dihedral angle in the I$\beta$ bulk which has an a-b-a-b-\dots layer structure. There are, however, large differences in the layers 1 and 3, and 2 and 4, respectively. We find $0.43:0.00:0.53:0.02$ for the $gg$ conformations, $0.23:0.65:0.15:0.31$ for the $gt$ conformations and $0.35:0.37:0.32:0.67$ for the $tg$ conformations. In the 8 layer supercell the differences between different flat and different rippled layers are much smaller, these data are included in table \ref{tbl:C6surf+water}; they are, nevertheless larger than those found for IIb, which has the same layer structure as I$\beta$.

\begin{table}
  \caption{ Number of hydrogen bonds within layers per cellobiose unit for the bulk and surface systems, excluding
intra-chain O3H$\cdots$O5 hydrogen bonds. The L stands for the layer in the slab. }
  \label{tbl: Hbond-surfsum}\centering
  \begin{tabular}{ccccc}
    \hline
    Phase & L  & crystal bulk & solid/vacuum & solid/water \\
    \hline
   Cellulose I$\alpha$
        & 1  & 2.10 & 2.28 & 1.94\\
	& 2  & 2.01 & 2.49 & 2.30\\
	& 3  & 1.99 & 2.30 & 2.35\\
   \hline
   Cellulose I$\beta$
               & 1  & 2.60 & 2.92 & 1.92 \\
		       & 2  & 3.51 & 3.34 & 3.19\\
		       & 3  & 2.63 & 2.48 & 2.49 \\
   \hline
   Cellulose IIa
          & 1  & 2.12 & 2.48 & 2.11 \\
		  & 2  & 3.01 & 2.41 & 2.60 \\
		  & 3  & 2.09 & 2.20 & 2.22  \\
   \hline
   Cellulose IIb
          & 1  & 2.26 & 2.68 & 2.26 \\
		  & 2  & 2.66 & 2.41 & 2.23 \\
		  & 3  & 2.36 & 2.20 & 2.35 \\
   \hline
  \end{tabular}
\end{table}

The HB networks in cellulose and at the interface are made from more than just the C6OH groups, so we calculated also
the average number of all intrachain and interchain HBs, except O3$\cdots$O5 HBs per cellobiose unit for each layer in
the bulk and interface systems (see Table \ref{tbl: Hbond-surfsum}).
The analysis did not allow to include intersheet hydrogen bonds, but the unchanged layer distances suggests that the
amount of interlayer hydrogen bonding is not strongly affected by interface reconstruction.

Again, we find for the four systems rather different trends: in the top layer of I$\alpha$, that shows the strongest
surface reconstruction, there is a moderate increase of the HBs when going from bulk to solid/vacuum, and a slightly
stronger decrease when going to  solid/water because the cellulose..water HBs are not counted. In the second and third
layer the number of HBs increases with respect to the bulk. Similar trends are found for the I$\beta$ system.

In the cellulose II systems we observe in the top layer an increase of the number of HBs when going from bulk to
solid/vacuum and a strong decrease when going from solid/vacuum to solid/water, there one has about as many HBs as in
the crystal layer. In layer 2 of both interface systems there is a decrease of the number of HBs with respect to the bulk,
in layer 3 the resemblance to the bulk is striking.

When a surface is created by cutting a slab out of the crystalline bulk, there will be hydroxy groups in the top layer
having no partner to form HBs between the layers (dangling HBs). During the reconstruction of the solid/vacuum interface
dangling HBs will be partially converted into inter-chain or maybe also inter-sheet HBs (which where not covered in the
analysis of table \ref{tbl: Hbond-surfsum}) in the top layer. The additional HBs, due to vacuum exposure one can directly
observe for Ialpha (100) in figure \ref{fgr:Ialpha-Reconstruction}. In the solid/water interface the former dangling HBs
and some of the inter-chain HBs in the top layer form HBs to the water slab. This yields a reduced average amount of HBs
within the first cellulose layer. We did not find, however, a significant amount of water molecules penetrating between
cellulose chains and diffusion of water molecules into the solid bulk; the formation of hydrated hydroxy groups at the
cellulose surface does not seem to be energetically sufficient, at least not in the simulation times used in this study.
This leaves us to conclude that pure water at ambient conditions is not able to cause swelling of cellulose
at microscopic timescales, but merely wetting.

\subsection*{\label{surf+water} Structure and dynamics of water at the interface}

\begin{figure}
\includegraphics[width=\textwidth]{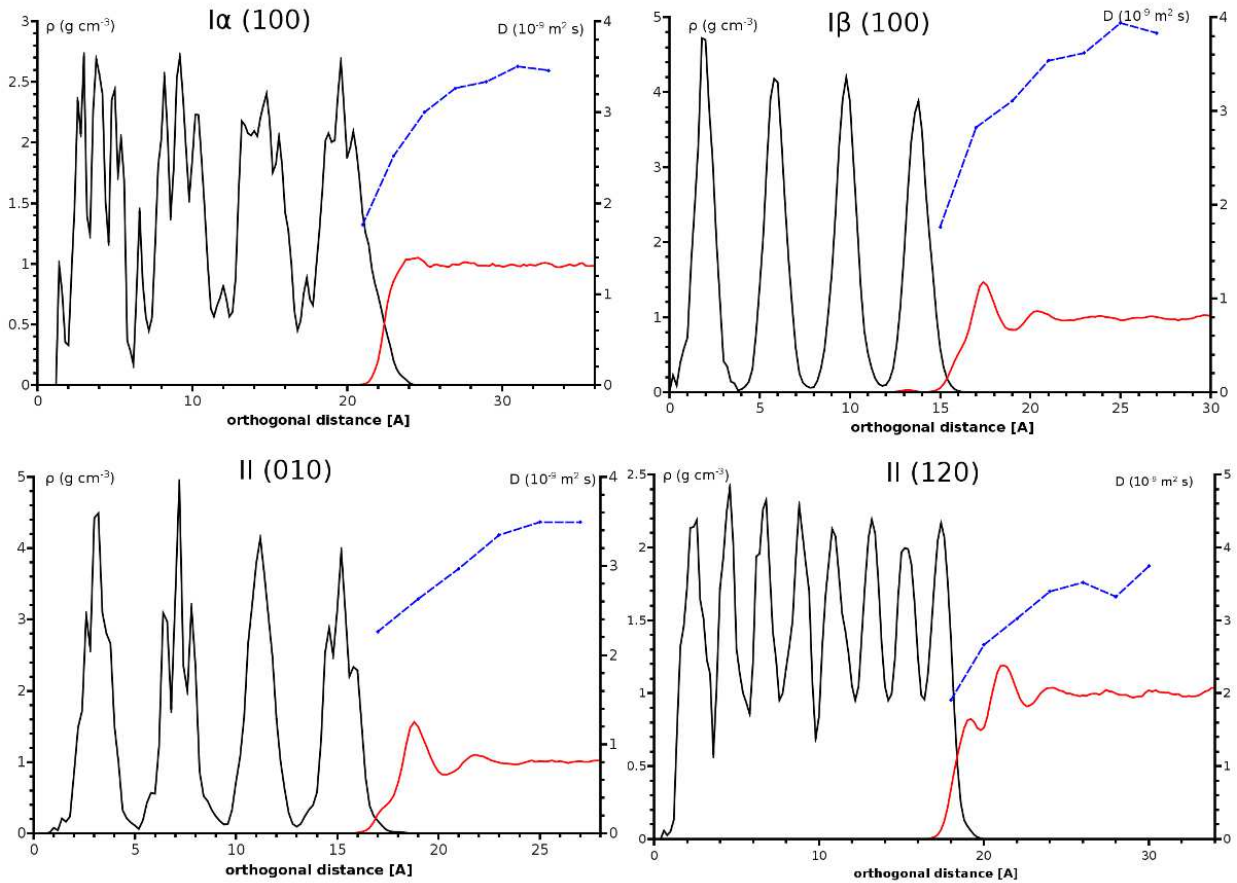}
\caption{\label{fgr: Ialphawater} Density profiles of cellulose-water interfaces, the cellulose profiles  (solid left) are for
I$\alpha$ (100), I$\beta$ (100), II (010) and II (120), right are the profiles (solid right). The   left scale in g$\;$cm$^{-3}$ is for the density profiles. Also shown is the diffusion coefficient of water molecules  as a function of the surfaces distance  (dashed) with the right scale  in 10$^{-9}\;$m$^2\,$s$^{-1}$. }
\end{figure}

To investigate the structure of the water slabs at the interfaces we calculated density profiles of the systems
orthogonal to the surface, which are shown in Figure \ref{fgr: Ialphawater}.
The black lines depict the density profiles of the cellulose, the red ones the density of the water.

The cellulose profiles reflect nicely the different layering in the four systems: fine structure is found when the
sheets are not parallel to the surface as for I$\alpha$ and IIa, the larger the angle between surface and sheet is,
the more structure is found in the profiles. For I$\beta$ and IIb, where the sheets are parallel to the surface,
we just see the peaks of the four and eight layers, respectively, in the systems.

The density profiles of water have an oscillatory form indicating an ordering in the water slab that is not typical
for bulk water, the highest density maximum is found near the interfaces. The deviations from the bulk density are strongest for the flat surfaces I$\beta$ (100) (40 percent) and II (010) (60 percent), and least
pronounced for the non-flat surfaces I$\alpha$ (100) (10 percent increase) and II (120)  (20 percent increase). The oscillations decay completely within 10$\;$\,\AA\, at the I$\alpha$ (100) surface the water bulk is undisturbed already after about 4 to 5$\;$\AA.

\begin{figure}
\includegraphics[width=\textwidth]{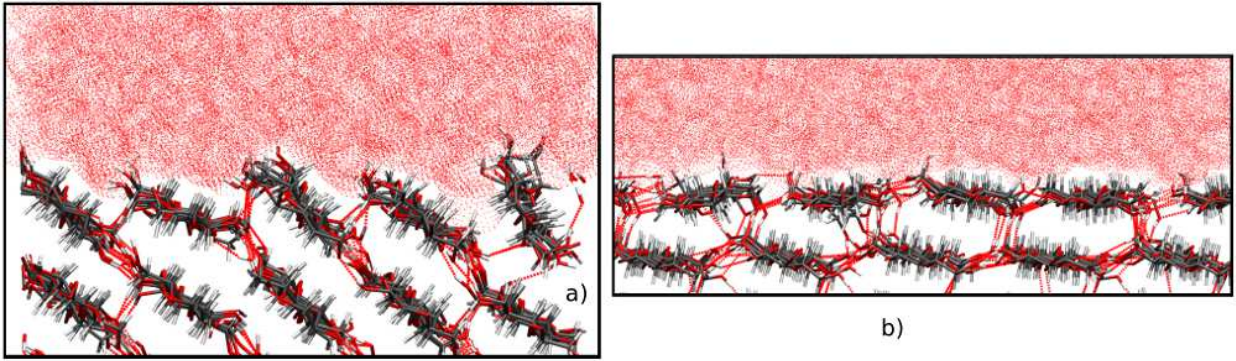}
\caption{\label{fgr: Waterarches} a,b base plane projection of a) the cellulose I$\alpha$ (100) and b) the cellulose I$\beta$ (100) slab in contact with water at room temperature and room pressure. Shown are the first layers of cellulose chains as sticks and the adsorbed water molecules in a continuous van der Waals surface representation. }
\end{figure}

When placing an undisturbed slab of water onto a cellulose slab, the water molecules rearrange to maximize interactions. We therefore see areas around the cellulose chains, where water molecules adsorb preferentially (around the terminal hydroxy groups) and areas where no hydrogen bonds can be formed, such as around the unpolar hydrogens bonded to the glucan framework. These areas show depletion of water molecules compared to the terminal hydroxy groups (cf. Figure \ref{fgr: Waterarches}). This densification and restructuring of water on the cellulose surfaces happens in a volume of a few \AA\ of vertical distance to the surface. Using a similar reasoning as \citep{Biermann01} one can relate the degree of hydrophilicity to the density profile: The more sharp the first peak is, corresponding to a dense layer of water adsorbed on the surface, the more hydrophilic is the surface. However, densification of a water slab above a cellulose surface can also be related to the regularity of the surface. Highly regular
surfaces, such as cellulose I$\beta$ (100) and cellulose II (010), allow a large amount of water molecules within a slab of a certain vertical spacing above the surface.  In the case of cellulose I$\alpha$ (and also cellulose II (120) the tilted and slightly irregular chains distribute this effect over a larger vertical range above the surface, although the amount of interacting water molecules may be similar, if not more. The densification of a water slab above a cellulose surface is therefore both, a function of the hydrophilicity and the regularity (roughness) of the surface. We therefore refrain from extrapolating solely from the density profile to the hydrophilicity of the surface, as has been done in literature \citep{Heiner98,Biermann01}, as long as we don't know in detail how the interaction energy at the water/surface interface depends on the surface morphology.  The surface structure of
cellulose II (120) surface with eight layers (see Figure \ref{fgr: Ialphawater}) allows interaction of water molecules not only with the top layer but also with the second layer, this is reflected in the shifted onset of the water profile which is at the same position as the maximum of the density of the cellulose top layer. Cellulose II (120) should therefore have a larger water accessible surface area and we hoped to observe initial solvation processes. No such processes were observed, however, within the simulation time. This  supports the finding that pure water and cellulose do not interact strongly enough to allow easy solubilization at ambient conditions.

\begin{table}
  \caption{Diffusion coefficients in m$\;$s$^{-2}$ for the first 2$\;$\,\AA\ thick layer of water in contact with
the different cellulose crystal surfaces at 300 K.}
  \label{tbl:diffusion}
  \begin{tabular}{cc}
    \hline
    Crystal surface & D$\times 10^9$ [m$\;$s$^{-2}$] \\
    \hline
    Cellulose I$\alpha$ (100) & 1.80(1)\\
    Cellulose I$\beta$ (100) & 2.11(1) \\
    Cellulose II (010) & 2.30(1) \\
    Cellulose II (120) & 1.90(1) \\
    \hline
  \end{tabular}
\end{table}
Another property that can be used to describe the structure of the water slab at the cellulose surface is the self diffusion
coefficient as a function of surface distance. The non-rigid TIP3P water model used in this study yields for bulk water at 300$\;$K a value of 4.0$\,10^9m\;s^{-2}$, which agrees well with our values at 10\,\AA, this value is about 50 percent larger than the experimental value of $2.4\times 10^{-9}\rm\,m^2\,s^{-1}$ (Refs. \citep{Harris1980,Holz2000}), but considerably better than the value that is obtained with the rigid TIP3P model (see Ref \citep{Vega2009}), although at considerably higher computational cost. Disregarding the absolute values of the diffusion constant, one can nevertheless obtain an interesting information on the cellulose/water interface: we find a decrease in the diffusion constant from bulk water to water at the interface of about
$1.6\times 10^{-9}\rm\,m^2\,s^{-1}$ and such a decrease corresponds, according to the data by Harris \textit{et al.}\citep{Harris1980}, to a decrease
of the water temperature of about 30 degrees. The reduction of the diffusion constant at the interface is caused by at least two factors: i) the reduced degrees of freedom of water motion, and ii) the formation of variably dense water regions (\lq arches\rq) at the cellulose/water interface of different stability. One could wonder if the shape of the graphs of the diffusion constant vs. distance from the surface reflects the differences in the water structures as
described by the density profiles; it is however not possible to show such a dependence, because the smallest
bin size used for the calculation of the self diffusion constant by integration of the velocity autocorrelation function was a slab of 2\,\AA, which is larger than the strongly fluctuating interface region. In any case, the reduced water mobility and the formation of water structures of different stability
at the cellulose surfaces, together with the change in hydrogen bonding patterns can be seen as a main ingredient of wetting of crystalline cellulose.
Our results are consistent with the results from a $^2$H NMR study\citep{Radloff96} who found three different types of water at the cellulose interface: a strongly bound, rigid water, which cannot crystallize, a highly mobile water, whose
motion is hindered due to anisotropic constraints, so that it can only undergo fast 180�
flips around their bisector axis, and finally the mobile water that can perform isotropic motions. This agrees also very well with the results from $^1$H and $^{13}$C solid state NMR results by Taylor \emph{et al.}\citep{Taylor08} that moisture adsorbed at cellulose surfaces has different correlation times.

As shown by Vega \emph{et al.} \citep{Vega2009}, the choice of the water model crucially influences the properties of liquid water and water steam. It should not be much of a surprise if the water model also influences the dynamical properties of interfaces. Dynamical observables might depend strongly on how balanced the force fields used for the cellulose and water description are.

\section{\label{conclusion} Conclusion }

Cellulose is a biological material of immense structural complexity. In this work we approached this material starting from idealized crystalline bulk structures. We  investigated the structure of different bulk crystals that allows a somewhat closed view. It is often claimed in literature that the attractive interaction between I$\alpha$ sheets is mainly dispersion interaction, whereas I$\beta$ and II show strong polar intersheet interactions; it should be clear that dispersion interaction is important for all kind of interfaces, irrespective of whether there are hydrogen bonds or not. Due to similar distances in our classical dynamical approach we always find a constant background of dispersion interaction for all crystal systems. Of course, one should know how much of the stabilization energy is dispersion energy and how much is, say, hydrogen bonding energy, but to do this quantum theoretical methods must be used. Work is in progress in our group to investigate the stabilization of
chains in sheets, or between sheets in bulk, and to calculate the contribution of dispersion interaction to stabilize these structures.

Knowledge of the reorganization of these bulk materials on interfaces is a basic prerequisite to understand more complex interactions that would involve ions, restructuring at non-standard conditions and, most interestingly, adsorption of complex nanostructured materials. When cellulose surface slabs are allowed to reconstruct, the degree of reconstruction depends strongly on the allomorph and the type of crystal plane that is chosen as slab surface. As a main result we find that within nanoseconds no significant mesoscopic structural changes, such as changes in layer distances, chain tilting or strong irregularities, occur. This might lead to the conclusion that cellulose surfaces in contact with vacuum or water show no restructuring at all. A detailed view on the molecular structure and the hydrogen bonding network reveals substantial restructuring, although only for the first few layers.

Adsorption of water on cellulose surfaces cannot be solely explained with the formation of hydrogen bonds between water molecules and oxygen containing groups at the surface, as in the bulk, dispersion interaction must be considered as well. And to find out, whether the interaction between glucan molecules, that is hydrogen bonds and dispersion interaction, are quantitatively correct described with force fields and water models, comparative quantum theoretical investigations are needed, specifically addressing the hydrogen bond strength and amount of dispersion interaction between the chains in cellulose. Such investigations are under way in our group  (Hoja, Grossar, Maurer and Sax, to be published). The results presented here point to different wettability of different cellulose surfaces, due to a different ratio of polar and unpolar moieties. This observation has already been made in recent studies\citep{Heiner98,Matthews06}. We can confirm those results and augment them with the observation that even
loosely packed high index surfaces, such as Cellulose II (120) do not allow significant solvation and swelling within a mesoscopic time frame. We further found hints that
water applied to surfaces that were optimized against vacuum can partially reverse the surface reconstruction, even though they did not insert between cellulose sheets, that were previously reconstructed against vacuum. Anyhow, we did not find that swelling of cellulose can be caused by pure water.

\begin{acknowledgments}
 The authors like to acknowledge financial support by the
European Union Framework Program 7 Project SURFUNCELL, Grant agreement No.: 214653, and the helpful comments of an unknown reviewer.
\end{acknowledgments}

%\bibliographystyle{ieeetr}
%\bibliography{cellusub}

\begin{thebibliography}{10}

\bibitem{Sullivan97}
A.~O'Sullivan, ``{Cellulose: the structure slowly unravels},'' {\em Cellulose},
  vol.~4, pp.~173--207, 1997.

\bibitem{Eichhorn10}
S.~J. Eichhorn, A.~Dufresne, M.~Aranguren, N.~E. Marcovich, J.~R. Capadona,
  S.~J. Rowan, C.~Weder, W.~Thielemans, M.~Roman, S.~Renneckar, W.~Gindl,
  S.~Veigel, J.~Keckes, H.~Yano, K.~Abe, M.~Nogi, A.~N. Nakagaito, A.~Mangalam,
  J.~Simonsen, A.~S. Benight, A.~Bismarck, L.~A. Berglund, and T.~Peijs,
  ``{Review: current international research into cellulose nanofibres and
  nanocomposites},'' {\em J. Mater. Sci.}, vol.~45, pp.~1--33, Sept. 2010.

\bibitem{Hornig08}
S.~Hornig and T.~Heinze, ``{Efficient approach to design stable
  water-dispersible nanoparticles of hydrophobic cellulose esters.},'' {\em
  Biomacromol.}, vol.~9, pp.~1487--1492, May 2008.

\bibitem{Kolpak76}
F.~J. Kolpak and J.~Blackwell, ``{Determination of the structure of cellulose
  II.},'' {\em Macromol.}, vol.~9, no.~2, pp.~273--278, 1976.

\bibitem{Wada2004}
M.~Wada, H.~Chanzy, Y.~Nishiyama, and P.~Langan, ``{Cellulose III Crystal
  Structure and Hydrogen Bonding by Synchrotron X-ray and Neutron Fiber
  Diffraction},'' {\em Macromol.}, vol.~37, no.~23, pp.~8548--8555, 2004.

\bibitem{Gardner74a}
K.~H. Gardner and J.~Blackwell, ``{The structure of native cellulose},'' {\em
  Biopolymers}, vol.~13, no.~10, pp.~1975--2001, 1974.

\bibitem{Gardner74b}
K.~H. Gardner and J.~Blackwell, ``{The hydrogen bonding in native cellulose},''
  {\em Biochim. Biophys. Acta}, vol.~343, no.~1, pp.~232--237, 1974.

\bibitem{Langan99}
P.~Langan, Y.~Nishiyama, and H.~Chanzy, ``{A Revised Structure and
  Hydrogen-Bonding System in Cellulose II from a Neutron Fiber Diffraction
  Analysis},'' {\em J. Am. Chem. Soc.}, vol.~121, pp.~9940--9946, Nov. 1999.

\bibitem{Nishiyama02}
Y.~Nishiyama, P.~Langan, and H.~Chanzy, ``{Crystal Structure and
  Hydrogen-Bonding System in Cellulose I$\beta$ from Synchrotron X-ray and
  Neutron Fiber Diffraction},'' {\em J. Am. Chem. Soc.}, vol.~124, no.~31,
  pp.~9074--9082, 2002.

\bibitem{Nishiyama03}
Y.~Nishiyama, J.~Sugiyama, H.~Chanzy, and P.~Langan, ``{Crystal structure and
  hydrogen bonding system in cellulose I(alpha) from synchrotron X-ray and
  neutron fiber diffraction.},'' {\em J. Am. Chem. Soc.}, vol.~125,
  pp.~14300--14306, Nov. 2003.

\bibitem{Perez96}
S.~Perez, M.~Kouwijzer, K.~Mazeau, and S.~Balling~Engelsen, ``{Modeling
  polysaccharides: Present status and challenges},'' {\em J. Mol. Graph.},
  vol.~14, pp.~307--321, Dec. 1996.

\bibitem{Kroon-Batenburg97}
L.~M.~J. Kroon-Batenburg and J.~Kroon, ``{The crystal and molecular structures
  of cellulose I and II},'' {\em Glycoconjugate J.}, vol.~14, pp.~677--690,
  Aug. 1997.

\bibitem{Mazeau03}
K.~Mazeau and L.~Heux, ``{Molecular Dynamics Simulations of Bulk Native
  Crystalline and Amorphous Structures of Cellulose},'' {\em J. Phys. Chem. B},
  vol.~107, pp.~2394--2403, Mar. 2003.

\bibitem{Vietor00}
R.~J. Vietor, K.~Mazeau, M.~Lakin, and S.~Perez, ``{A priori crystal structure
  prediction of native celluloses},'' {\em Biopolymers}, vol.~54, no.~5,
  pp.~342--354, 2000.

\bibitem{Heiner95}
A.~P. Heiner, J.~Sugiyama, and O.~Teleman, ``{Crystalline cellulose I [alpha]
  and I [beta] studied by molecular dynamics simulation},'' {\em Carbohydr.
  Res.}, vol.~273, pp.~207--223, Aug. 1995.

\bibitem{Bergenstrahle07}
M.~Bergenstrahle, L.~A. Berglund, and K.~Mazeau, ``{Thermal response in
  crystalline Ibeta cellulose: a molecular dynamics study.},'' {\em J. Phys.
  Chem. B}, vol.~111, pp.~9138--9145, Aug. 2007.

\bibitem{Kroon-Batenburg96}
L.~M.~J. Kroon-Batenburg, B.~Bouma, and J.~Kroon, ``{Stability of Cellulose
  Structures Studied by MD Simulations. Could Mercerized Cellulose II Be
  Parallel?},'' {\em Macromol.}, vol.~29, pp.~5695--5699, Jan. 1996.

\bibitem{Hardy96}
B.~Hardy and A.~Sarko, ``{Molecular dynamics simulations and diffraction-based
  analysis of the native cellulose fibre: structural modelling of the
  I-$\alpha$ and I-$\beta$ phases and their interconversion},'' {\em Polymer},
  vol.~37, pp.~1833--1839, May 1996.

\bibitem{Eichhorn05}
S.~J. Eichhorn, R.~J. Young, and G.~R. Davies, ``{Modeling crystal and
  molecular deformation in regenerated cellulose fibers.},'' {\em
  Biomacromol.}, vol.~6, no.~1, pp.~507--513, 2005.

\bibitem{Matthews06}
J.~F. Matthews, C.~E. Skopec, P.~E. Mason, P.~Zuccato, R.~W. Torget,
  J.~Sugiyama, M.~E. Himmel, and J.~W. Brady, ``{Computer simulation studies of
  microcrystalline cellulose I beta},'' {\em Carbohydr. Res.}, vol.~341,
  pp.~138--152, 2006.

\bibitem{Yui06}
T.~Yui, S.~Nishimura, S.~Akiba, and S.~Hayashi, ``{Swelling behavior of the
  cellulose I beta crystal models by molecular dynamics},'' {\em Carbohydr.
  Res.}, vol.~341, pp.~2521--2530, Nov. 2006.

\bibitem{Yui07}
T.~Yui and S.~Hayashi, ``{Molecular dynamics simulations of solvated crystal
  models of cellulose I-alpha and IIII},'' {\em Biomacromol.}, vol.~8,
  pp.~817--824, 2007.

\bibitem{Heiner97}
A.~P. Heiner and O.~Teleman, ``{Interface between Monoclinic Crystalline
  Cellulose and Water: Breakdown of the Odd/Even Duplicity},'' {\em Langmuir},
  vol.~13, pp.~511--518, Feb. 1997.

\bibitem{Biermann01}
O.~Biermann, E.~H\"{a}dicke, S.~Koltzenburg, and F.~M\"{u}ller-Plathe,
  ``{Hydrophilicity and Lipophilicity of Cellulose Crystal Surfaces},'' {\em
  Angew. Chem. Int. Ed.}, vol.~40, no.~20, pp.~3822--3825, 2001.

\bibitem{Mazeau2008}
K.~Mazeau and A.~Rivet, ``{Wetting the (110) and (100) surfaces of Ibeta
  cellulose studied by molecular dynamics.},'' {\em Biomacromol.}, vol.~9,
  pp.~1352--4, Apr. 2008.

\bibitem{Bergenstrahle08}
M.~Bergenstrahle, K.~Mazeau, and L.~Berglund, ``{Molecular modeling of
  interfaces between cellulose crystals and surrounding molecules: Effects of
  caprolactone surface grafting},'' {\em Eur. Polym. J.}, vol.~44,
  pp.~3662--3669, Nov. 2008.

\bibitem{Mazeau02}
K.~Mazeau and C.~Vergelati, ``{Atomistic Modeling of the Adsorption of
  Benzophenone onto Cellulosic Surfaces},'' {\em Langmuir}, vol.~18,
  pp.~1919--1927, Mar. 2002.

\bibitem{Kontturi11}
E.~Kontturi, M.~Suchy, P.~Penttil\"a, B.~Jean, K.~Pirkkalainen, M.~Torkkeli,
  and R.~Serimaa, ``{Amorphous characteristics of an ultrathin cellulose
  film.},'' {\em Biomacromol.}, vol.~12, no.~2, pp.~770--777, 2011.

\bibitem{Aulin09}
C.~Aulin, S.~Ahola, P.~Josefsson, T.~Nishino, Y.~Hirose, M.~Osterberg, and
  L.~Wagberg, ``{Nanoscale cellulose films with different crystallinities and
  mesostructures--their surface properties and interaction with water.},'' {\em
  Langmuir}, vol.~25, no.~13, pp.~7675--7685, 2009.

\bibitem{Smith05}
W.~Smith, I.~T. Todorov, and M.~Leslie, ``{The DL\_POLY molecular dynamics
  package},'' {\em Z. Kristallogr.}, vol.~220, no.~5-6, pp.~563--566, 2005.

\bibitem{Kirschner08}
K.~N. Kirschner, A.~B. Yongye, S.~M. Tschampel, J.~Gonzalez-Outeirino, C.~R.
  Daniels, B.~L. Foley, and R.~J. Woods, ``{GLYCAM06: a generalizable
  biomolecular force field. Carbohydrates.},'' {\em J. Comput. Chem.}, vol.~29,
  no.~4, pp.~622--655, 2008.

\bibitem{Jorgensen83}
W.~L. Jorgensen, J.~Chandrasekhar, J.~D. Madura, R.~W. Impey, and M.~L. Klein,
  ``{Comparison of simple potential functions for simulating liquid water},''
  {\em J. Chem. Phys.}, vol.~79, no.~2, pp.~926--935, 1983.

\bibitem{Case05}
D.~A. Case, T.~E. Cheatham, T.~Darden, H.~Gohlke, R.~Luo, K.~M. Merz,
  A.~Onufriev, C.~Simmerling, B.~Wang, and R.~J. Woods, ``{The Amber
  biomolecular simulation programs.},'' {\em J. Comput. Chem.}, vol.~26,
  no.~16, pp.~1668--1688, 2005.

\bibitem{Sherwood03}
P.~Sherwood, A.~de~Vries, M.~Guest, G.~Schreckenbach, C.~Catlow, S.~French,
  A.~Sokol, S.~Bromley, W.~Thiel, A.~Turner, S.~Billeter, F.~Terstegen,
  S.~Thiel, S.~Kendrick, J.~Kendrick, S.~Rogers, J.~Casci, M.~Watson, F.~King,
  E.~Karlsen, M.~Sjovoll, A.~Fahmi, A.~Sch\"afer, and L.~C, ``{QUASI: A general
  purpose implementation of the QM/MM approach and its application to problems
  in catalysis},'' {\em J. Mol. Struct.: THEOCHEM}, vol.~632, no.~1-3,
  pp.~1--28, 2003.

\bibitem{Youngs10}
T.~G.~A. Youngs, ``{Aten - An application for the creation, editing, and
  visualization of coordinates for glasses, liquids, crystals, and
  molecules},'' {\em J. Comput. Chem.}, vol.~31, no.~3, pp.~639--648, 2010.

\bibitem{Humphrey96}
W.~Humphrey, A.~Dalke, and K.~Schulten, ``{VMD -- Visual Molecular Dynamics},''
  {\em J. Molec. Graphics}, vol.~14, pp.~33--38, 1996.

\bibitem{Martinez09}
L.~Martinez, R.~Andrade, E.~G. Birgin, and J.~M. Martinez, ``{PACKMOL: A
  Package for Building Initial Configurations for Molecular Dynamics
  Simulations},'' {\em J. Comput. Chem.}, vol.~30, pp.~2157--2164, 2009.

\bibitem{Zhang11}
Q.~Zhang, V.~Bulone, H.~\AA~gren, and Y.~Tu, ``{A molecular dynamics study of
  the thermal response of crystalline cellulose Ibeta},'' {\em Cellulose},
  vol.~18, no.~2, pp.~207--221, 2011.

\bibitem{Chen12}
P.~Chen, Y.~Nishiyama, and K.~Mazeau, ``{Torsional Entropy at the Origin of the
  Reversible Temperature-Induced Phase Transition of Cellulose.},'' {\em
  Macromol.}, vol.~45, no.~1, pp.~362--368, 2012.

\bibitem{Heiner98}
A.~P. Heiner, L.~Kuutti, and O.~Teleman, ``{Comparison of the interface between
  water and four surfaces of native crystalline cellulose by molecular dynamics
  simulations},'' {\em Carbohydr. Res.}, vol.~306, pp.~205--220, 1998.

\bibitem{Wada2004_1}
M.~Wada, L.~Heux, and J.~Sugiyama, ``{Polymorphism of Cellulose I Family:
  Reinvestigation of Cellulose IV$_I$},'' {\em Biomacromol.}, vol.~5,
  pp.~1385--1391, 2004.

\bibitem{Yamamoto89}
H.~Yamamoto, F.~Horii, and H.~Odani, ``{Structural-Changes of native cellulose
  crystals induced by annealing in aqueous alkaline and acidic solutions at
  high-temperatures},'' {\em Macromol.}, vol.~22, no.~10, pp.~4130--4132, 1989.

\bibitem{Degarmo03}
E.~P. Degarmo, J.~T. Black, and R.~A. Kohser, {\em {Materials and Processes in
  Manufacturing}}.
\newblock Wiley, 9th~ed., 2003.

\bibitem{Mazeau11}
K.~Mazeau, ``{On the external morphology of native cellulose microfibrils},''
  {\em Carbohydr. Polym.}, vol.~84, no.~4, pp.~524--532, 2011.

\bibitem{Newman04}
R.~H. Newman and T.~C. Davidson, ``{Molecular conformations at the
  cellulose--water interface},'' {\em Cellulose}, vol.~11, pp.~23--32, 2004.

\bibitem{Baker00}
A.~A. Baker, W.~Helbert, J.~Sugiyama, and M.~J. Miles, ``{New Insight into
  Cellulose Structure by Atomic Force Microscopy Shows the I$\alpha$ Crystal
  Phase at Near-Atomic Resolution},'' {\em Biophysical J.}, vol.~79,
  pp.~1139--1145, 2000.

\bibitem{Harris1980}
K.~R. Harris and L.~A. Woolf, ``{Pressure and temperature dependence of the
  self diffusion coefficient of water and oxygen-18 water},'' {\em J. Chem.
  Soc., Faraday Trans. 1 F}, vol.~76, pp.~377--385, 1980.

\bibitem{Holz2000}
M.~Holz, S.~Heil, and A.~Sacco, ``{Temperature-dependent self-diffusion
  coefficients of water and six selected molecular liquids for calibration in
  accurate 1H NMR PFG measurements},'' {\em Phys. Chem. Chem. Phys.}, vol.~2,
  pp.~4740--4742, 2000.

\bibitem{Vega2009}
C.~Vega, J.~Abascal, M.~Conde, and J.~Aragones, ``{What ice can teach us about
  water interactions: a critical comparison of the performance of different
  water models},'' {\em Faraday Discuss.}, vol.~141, pp.~251--276, 2009.

\bibitem{Radloff96}
D.~Radloff, C.~Boeffel, and H.~W. Spiess, ``{Cellulose and Cellulose/Poly(vinyl
  alcohol) Blends. 2. Water Organization Revealed by Solid-State NMR
  Spectroscopy},'' {\em Macromol.}, vol.~29, no.~5, pp.~1528--1534, 1996.

\bibitem{Taylor08}
R.~E. Taylor, A.~D. French, G.~R. Gamble, D.~S. Himmelsbach, R.~D. Stipanovic,
  D.~P. Thibodeaux, P.~J. Wakelyn, and C.~Dybowski, ``{$^1$H and $^{13}$C solid
  state NMR of Gossypium barbadense (Pima) cotton},'' {\em J. Mol. Struct.},
  vol.~878, pp.~177--184, 2008.

\end{thebibliography}

\end{document}